# COMPUTATIONAL METHODS for REQUIRED MATERIAL'S COMPOSITION and STRUCTURE.

Konstantin Borodianskiy and Michael Zinigrad

*Laboratory for Metallic and Ceramic Coatings and Nanotechnology, Ariel University Center of Samaria, Ariel, 40700, Israel.*

**Abstract**

The mathematical modeling of real high temperature processes, such as welding, casting, joining technologies, based on a physicochemical analysis of the interaction between the phases is in the focus of the present work. The model is based on the fundamental equations of thermodynamics and kinetics of high-temperature metallurgical reactions and factors which take into account the thermal and hydrodynamics conditions of the real process.

The model can be used to predict the chemical composition of the metal matrix, as well as the quantitative and qualitative composition of the strengthening carbide phases formed during primary and secondary crystallization processes.

The correctness of the developed model calculation was experimentally examined in a real technological problem, *i.e.*, the development of a new flux-cored wire for forming a build-up layer with certain shock-abrasion resistance properties. The inverse problem of flux-cored wire computation was solved, and can provide us with the required chemical composition of a build-up metal and its structure. These required properties are achieved by an austenite-martensite matrix structure with a low percentage of carbides phases uniformly distributed therein.

The build-up metal was prepared and tested to determine its chemical composition, structure and mechanical properties. The obtained chemical composition structure and mechanical properties were compared with the experimentally results. It was found that the experimental results confirm the adequacy of the computer calculations using this model.

## 1. State of the art.

One of the most important and complicated problems in modern industry is to obtaining materials with the required chemical composition, structure and mechanical and physical properties. Solving this problem involves great deal of time and expense, and the results obtained might be far from the optimal solution.

The development of computer technology and its accessibility have made it possible to solve problems for which there were previously unknown solutions or these methods were so tedious that they proved to be unsuitable for practical application.

There are some works where models were developed to predict the chemical composition [1, 2] and the structure [3-20] of the required deposit metal during high temperature processes, such as welding, joining, and build-up processes. These complicated models include the physical and chemical parameters of solid, liquid and gas phases, phase transition parameters, hydrodynamics' parameters, *etc*.

It is also necessary to be aware of the material's phase structure, which has a significant influence on its final properties. Predicting the phase-structure composition of a metal has been the subject of numerous papers [21-27] that included graphical representations of the phase-structure composition of the metal as a function of its chemical composition, as well as computational methods for determining the percentage of its phase.

Such a method is poorly suited to complex systems and processes described by systems of equations. In the case of mathematical modeling, the process is studied on a mathematical model using a computer, and not on a physical object. The input parameters of the mathematical model are fed into the computer, and the computer supplies the output parameters calculated in the process. The first stage in the mathematical modeling of physicochemical systems is generally the construction of thermodynamic models. This stage is very important both for ascertaining the fundamental possibility of the combined occurrence of particular chemical processes and for listing the most important thermodynamic characteristics.

In recent years mathematical modeling has been applied not only to the investigation of theoretical aspects of physicochemical processes, but also to the analysis of real technologies.

The areas of the prediction and optimization of the composition and properties of materials obtained in different technological processes are especially promising. Some of the results were obtained from the modeling of the process of the formation of a weld pool, from modeling of weld metal transformations. Important results were obtained from the studies of the physical and chemical parameters of high temperature processes, such as welding or casting, and development of kinetic model of alloy transfer. By determining the chemical composition of the weld metal researchers have developed the kinetic model [3-9]. Based on this model, the authors described the transfer of alloying elements between the slag, which is the residue left on a weld from the flux consists mostly of mixed metal oxides, sulfides and nitrides, and the metal during arc welding. The model takes into consideration the practical weld process parameters such as voltage, current, travel speed, and weld preparation geometry, and it was experimentally tested.

## 2. Structural composition.

Structural approach is only a method development tool, *i.e.*, a means to structurize a problem, to establish connections and the order of priorities, to structurize data, *etc.*, using structural analysis.

A brief review of the major stages of structural analysis of the welding materials design problems is presented below:
1. Determining the composition (structure) of the design object subject domain.
2. Establishing functional relationship between the design object and the subject domain elements (direct and reverse connections).
3. Establishing connections between the subject domain and the design tool (expert system).
4. Determining the operational algorithm structure and the subject domain representation method for the design tool.
5. Design stages structurization. Setting priorities.
6. Establishing functional relationship between the design stages, tools and design object.
7. Determining input and output parameters of the design tool.
    This is schematically shown in Fig. 1.

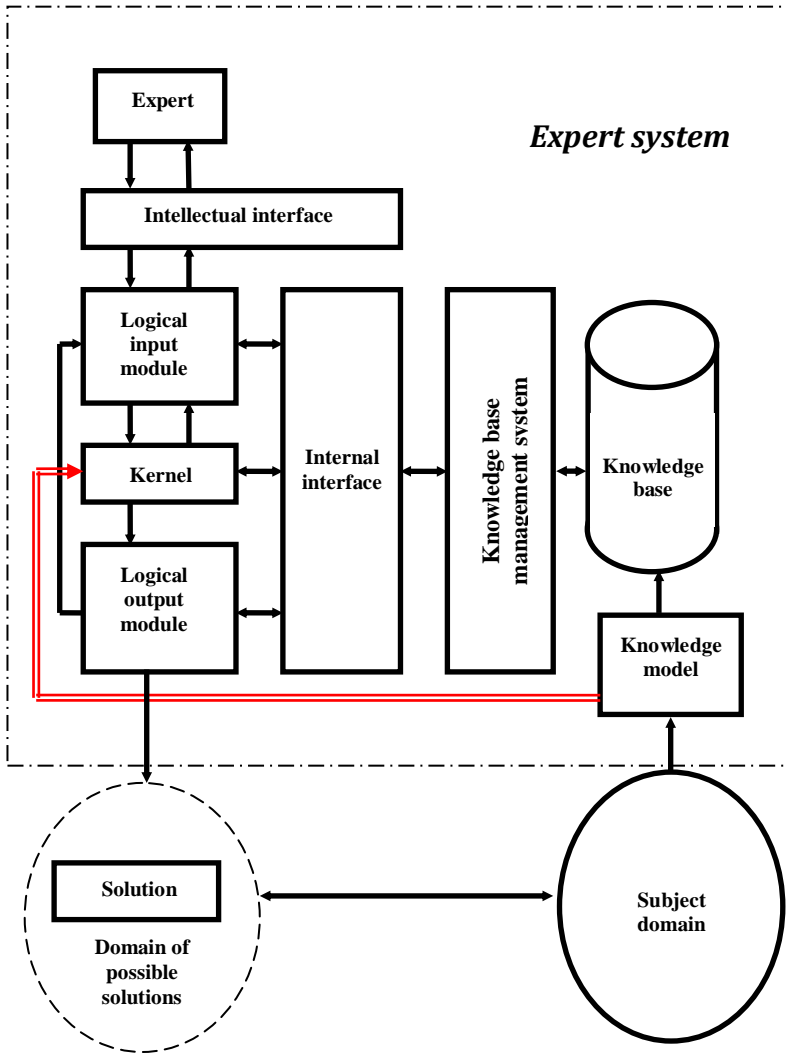

**Fig. 1.** Simplified flow-chart of the expert system.

### 3. Mathematical modeling of phase interaction in real technological process.

A detailed analysis of a welded joint and its interaction with the environment are presented in the followed part. It was shown that an effective method for developing a new welding material involves solving the inverse problem of finding the formula of the material as a function of the service characteristics of the weld metal. The most important problems for the new methodology in the area of determining the electrode formula of a new welding material include devising a model of the required structure of the weld metal under service conditions and calculating the primary structure and chemical composition of the weld deposit. The chemical composition of the weld metal is determined by the initial chemical composition of the welding material and the base metal and by the nature of the physiochemical processes accompanying the interaction between the molten metal and slag.

Prediction of the chemical composition of a weld deposit and, consequently, determine its mechanical properties, is based on a kinetic analysis of the simultaneous diffusion-controlled reactions that occur between the molten metal and slag [5]. The mutual influence of the reactions and the diffusion of all the reactants in the metal and slag are also taken into account.

The analysis of the kinetics and mechanism of individually occurring reactions does not present any special difficulties at the present time and that, as a rule, its results faithfully describe the real processes. A kinetic analysis of the interaction of multi-component metallic and slag melts with consideration of the mutual influence of reactions taking place in parallel is considerably more complicated. The theoretical basis of the method consists of two assumptions:

- under diffusion-controlled conditions the concentration ratio at the phase boundary for each reaction is close to the equilibrium value;
- the rate of transfer of the reactants to the phase boundary or away from it is proportional to the difference between their concentrations in the bulk and on the boundary of the metallic and oxide melts.

The oxidation of elements in a metallic melt can be represented by the reaction.

$$\frac{n}{m}[E_i] + (FeO) \rightarrow \frac{1}{m}(E_{i_n}O_m) + Fe \qquad (1)$$

where $E_i$ denotes the elements dissolved in the molten metal (Mn, Si, W, Mo, V, *etc.*), and $E_{in}O_m$ denotes the oxides in the molten slag.

A calculation of the rates of reactions of type (1) for each element does not present any difficulties. However, a separate analysis of each reaction does not correspond to the real industrial processes occurring in the weld pool. The mutual influence of both the components of the interacting molten phases and the heterogeneous reactions taking place in these complex systems must be considered. Within the approach developed, the rate $v_{Ei}$ of mass transfer of any element (mol/cm$^2$s) for reactions of type (1) for all the metal components with consideration of their mutual influence are defined by the expression:

$$V_{Ei} = \frac{x^m - K_{Ei}^m \times \frac{(E_{in}O_m)}{[Ei]}}{\frac{x^m}{v_{Ei}^{\lim}} + \frac{K_{Ei}^m \times (E_{in}O_m)}{[Ei] \times v_{E_{in}O_m}^{\lim}}} \qquad (2),$$

at n=1.

$$V_{Ei} = v_{Ei}^{\lim}\left[1 + b_i \times \frac{v_{Ei}^{\lim}}{4v_{Ei_2O_m}^{\lim}} - \sqrt{\left(1 + b_i \times \frac{v_{Ei}^{\lim}}{4v_{Ei_2O_m}^{\lim}}\right)^2 - 1 + b_i}\right] (3),$$

where *b* is defined by the equation:

$$b_i = \frac{K_{Ei}^m (Ei_2O_m)}{x^m [E_i]^2} \quad (4),$$

at n=2.

Here $K_{Ei}$ is the equilibrium constant of reaction (1) for the *i-th* component of the molten metal, and $n$ and $m$ are stoichiometric coefficients, $x$ is the ratio between the concentration of iron oxide in the slag and the concentration of iron in the molten metal on the boundary between the interacting phases:

$$x = \frac{(FeO)'}{[Fe]'} \quad (5),$$

$v^{lim}_{Ei}$ and $v^{lim}_{EinOm}$ are the limiting diffusion fluxes of the components (j) of the molten metal or slag phases, [Ei] and ($E_{in}O_m$) are the initial concentrations (wt.%) of the elements and oxides in the molten phases, respectively calculated by:

$$v_j^{lim} = \beta \times D_j^{1/2} \times C_j \quad (6)$$

where $\beta$ is the mass-transfer coefficient (cm/s), $D_j$ is a diffusion coefficient (cm²/s), and $C_j$ is a reagent's concentration at the phase boundary (mol/cm³).

The rate $V_{FeO}$ of mass transfer of iron monoxide (mol/cm²s) which is the second reagent in reactions of type (1) defined by the expression:

$$V_{FeO} = \frac{\frac{(FeO)}{[Fe]} - x}{\frac{x}{v_{Fe}^{lim}} + \frac{(FeO)}{[Fe] \times v_{FeO}^{lim}}} \quad (7).$$

It follows from the stoichiometry of the reaction (1) that:

$$V_{FeO} = \sum_1^k V_i = \Sigma \frac{m}{n} V_{Ei} \quad (8).$$

Having substituted $V_{FeO}$ and $V_{Ei}$ in expression (8), we will have an equation with one unknown—$x$. Having found $x$ from (2) and (3) we can find $V_{Ei}$.

The scheme of the analyzed technological process of fusion welding process is presented in Fig. 2.

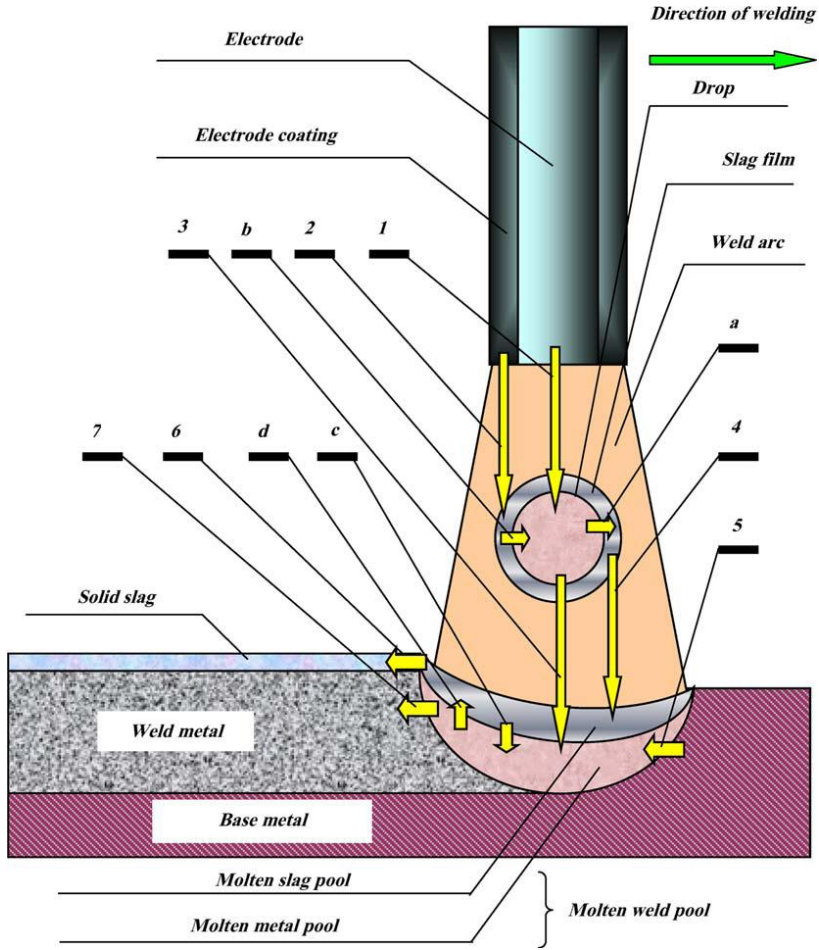

**Fig. 2.** Scheme of the fusion welding process.

On the scheme, figures denote the direction of material transfer, and letters denote the interaction of phases:
- melting of the electrode bare and formation of a drop (1);
- melting of the electrode coating and formation of slag film over the drop (2);
- transfer of the drop metal (which has reacted with slag film at the stage of transfer) to the metal pool (3);
- transfer of the slag film (which has reacted with the drop metal at the stage of transfer) to the slag pool (4);
- melting of base metal (5);
- crystallization of the slag pool (6);
- crystallization of the metal pool (7);
- a, b—redox reaction at slag–metal boundary in a welding drop;
- c, d—redox reaction at slag–metal boundary in a welding pool.

The final composition of the drop in general case is determined by the concentrations in each of the powdered components in the flux of the flux-cored wire or in the electrode coating $[Ei]^l_{pd}$ and, accordingly, by the melting rates of these components $v^l_{pd}$, the concentration of each element in the metal sheath or in the electrode bare $[Ei]_b$ and its melting rate $v_b$, as well as by the rates of passage of the elements through the interface $V_{Ei}$ of the drop and the slag film on its surface, which can be calculated in accordance with the methods described above. The values of $v^l_{pd}$ and $v_b$ are found from empirical relations as functions of the technological parameters of the process. Thus, the concentration of the *i-th* element in the drop at any moment in time *s*, can be calculated from the equation:

$$[Ei]^\tau = \frac{v_b \cdot [Ei]_b \cdot d\tau + \sum_{l=1}^{L} v^l_{pd} \cdot [Ei]^l_{pd} \cdot d\tau + 100 \cdot M_{Ei} \cdot A_d \cdot V_{Ei} \cdot d\tau}{m^\tau_d} \quad (9),$$

where $A_d$ is the surface area of the liquid drop, $l$ labels the type of powder, $L$ is the number of types, $M_{Ei}$ molar mass of the *i-th* element, and $m^\tau_d$ is the mass of the metal drop at the time *s*.

The final drop composition thus calculated $[Ei]_d$ is used to calculate the concentration of the *i-th* element in the weld pool at any time *s*. The composition of the pool and therefore the composition of the weld metal are determined by the concentrations of the elements in the liquid drop $[Ei]_d$ and accordingly by the rate of descent of the drops into the liquid pool $v_d$, the concentration of each element in the base metal $[Ei]_{bm}$, the melting rate of the base metal $v_{bm}$ and, as in the case of the drop stage, by the rate of passage of each element through the interface between the metal and slag pools.

In accordance with the foregoing, the expression for calculating the final composition of element *i* in the crystallized metal can be written in the form:

$$[Ei]^\tau = \frac{v_d \cdot [Ei]_d \cdot d\tau + v_{bm} \cdot [Ei]_{bm} \cdot d\tau + 100 \cdot M_{Ei} \cdot A_p \cdot \int_0^\tau V_{Ei} \, d\tau}{m^\tau_p} \quad (10)$$

where $A_p$ is the interfacial interaction area between the metal and the slag, and $m^\tau_p$ is the mass of the weld pool at the time *s*.

Thus, the proposed method can be used to find the chemical composition of the molten metal in the weld pool, *i.e.*, of the metal in a welded joint. This chemical composition is the starting point for determining the quantitative and qualitative composition of the phases of the weld deposit.

## 4. Physicochemical analysis of primary crystallization and carbide formation.

The subsequent transformations of the molten metal are associated with the primary and secondary crystallization processes, *i.e.*, the phase transformations in the multi-component alloy.

After determining the chemical composition of the *i-th* component in the solid phase at the crystallization time *τ*, we should determine its distribution between the austenite and the carbide phases that have formed at the primary crystallization process.

Let us use the chemical composition of the liquid molten metal in the weld pool as a starting point for examining the primary crystallization process. As we know from the theory of welding processes, crystallization of the weld pool proceeds under highly non-equilibrium conditions in the absence of convective stirring of the metal in the "tail" of the weld pool, *i.e.*, at the crystallization front. Therefore, the process of distributing the components between the liquid and solid phases is controlled only by diffusion. Another important factor that determines the distribution of the components is the concentration buildup occurring at the crystallization front. These factors produce concentration-induced supercooling, which, together with thermal supercooling, is responsible for the cyclic character of weld pool crystallization and the chemical nonuniformity of the crystallized weld metal. At any moment during crystallization of the weld pool, the amount of the *i-th* component that has passed from the liquid phase into the solid phase can be defined as:

$$Ei^s = Ei^0[1-(1-K_{eff})\exp(-\frac{L^t v_{cryst}}{D_i^l})] \qquad (11)$$

where $Ei^s$ is the concentration of the *i-th* component in the solid phase at the crystallization time τ, $Ei^0$ is the initial mean concentration of the *i-th* component in the molten phase, $K_{eff}$ is the effective distribution coefficient, $L^t$ is the distance from the crystallization starting point (the length of the crystallite at the crystallization time τ), $v_{cryst}$ is the crystallization rate, and $Di^l$ is the diffusion coefficient of the *i-th* component in the molten phase.

After determining the concentration of the *i-th* component in the solid phase at the crystallization time *t*, we still cannot determine its distribution between the austenite and the carbide phases that have formed at the crystallization process. The factors that influence carbide formation can be divided into two groups:
- physicochemical factors, which directly determine the nature of the carbide-formation process.
- technological factors, which indirectly influence the carbide-formation process by altering the physicochemical factors parameters.

In our work, the principles governing carbide formation in an alloyed iron-carbon weld deposit were formulated on the basis of a detailed physicochemical analysis of the formation of primary carbides as compounds of carbon with *d* metals according to the quantum-chemical theories of the electronic structure of *d* metals and primary carbides. The carbide forming reaction can be described as follows:

$$xEi + yC \rightarrow (Ei)_x C_y \qquad (12).$$

According to these principles, the amount of carbon that is used to form the carbide of the *i-th* metal is proportional to the atomic radius of the metal ($R_i$) and is inversely proportional to the number of electrons in the *d* sublevel of the metal ($d_i$). We introduce the concept of the absolute Carbide Forming Ability (CFA) tendency of the *i-th d* metal ($\Theta_i$) as the ratio:

$$\Theta_i = \frac{R_i}{d_i} \qquad (13).$$

It follows from an analysis of (12) that the carbide-forming tendency increases along the series consisting of: Fe, Mn, Cr, Mo, W, Nb, V, Ta, Ti, Zr, and Hf, in good agreement with the results in [24, 25]. The distribution of the alloying elements and carbon between the liquid and the solid phases is given by (11). Diffusionless decomposition of the supersaturated solid solution to austenite and carbide phases occurs during crystallization. The amount of carbon bound by any carbide-forming element is determined by the stoichiometry of the compound ($Me_xC_y$) and can be found from the following expression:

$$E_{Ci}^{(c)} = Ei^{(c)} \frac{yA_C}{xA_i} \quad (14),$$

where $x$ and $y$ are stoichiometric coefficients, $A_C$ and $A_i$ are the atomic weights of carbon and the carbide-forming element, respectively, and $E_i^{(c)}$ is the concentration of the carbide-forming element in the carbide phase. For primary carbides, the value of $x$ is always equal to 1, and $y$ takes values from 0.4 to 1.0, depending on the homogeneity region of the respective carbide. It is logical to assume that only the portion of the alloying elements and carbon that cannot be dissolved in austenite at the respective temperature is used for carbide formation:

$$E_C^{(c)t_k} = E_C^{(s)t_k} - E_C^{(\lim)t_k} \quad (15)$$

where $E_i^{(c)tk}$ is the concentration of carbon that is not dissolved in austenite, $E_i^{(s)tk}$ is the carbon concentration given by (13) at the crystallization time, and $E_i^{(\lim)tk}$ is the solubility limit of carbon in austenite at the respective crystallization temperature at the time $t_k$. The distribution of carbon between the carbide phases and the alloy will be proportional to the relative carbide-forming tendency of the respective transition element

$$\frac{\Theta_i}{\sum_{i=1}^{l} \Theta_i}$$

and its concentration in the alloy $a_i$. It is now clear that the proportionality factor for the $i$-th carbide-forming element is:

$$\eta_i = \frac{1}{2}\left(\frac{a_i}{100} + \frac{\Theta_i}{\sum_{i=1}^{n} \Theta_i}\right) \quad (16).$$

Then the concentration of the $i$-th carbide-forming element bound in the corresponding carbide phase at the time $t_{kt}$ can be defined as (wt.%):

$$Ei^{(c)t_k} = \eta_i \cdot E_C^{(c)t_k} \frac{xA_i}{yA_C} \quad (17),$$

and the concentration of the $i$-th carbide-forming element dissolved in austenite at the time $t_k$ determined by (wt.%):

$$Ei^{(a)t_k} = Ei^{(s)t_k} - Ei^{(c)t_k} \quad (18).$$

The concentration of carbides formed at the time $t_k$ (wt.%) is the sum of the carbon concentration and the total concentration of the carbide-forming elements that have participated in carbide formation:

$$Q_k^{t_k} = E_C^{(c)t_k} + \sum_{i=1}^{l} Ei^{(c)t_k} \quad (19).$$

Then the austenite content (wt.%) is:

$$S^{(a)} = 100\% - Q^I \qquad (20)$$

The mean concentrations (wt.%) of carbon and the alloying elements in the austenite phase can be found, respectively, as:

$$E_C^{(a)} = \frac{\sum_{k=1}^{z} E_C^{t_k}}{zS^{(a)}} 100\% \qquad (20),$$

$$Ei^{(a)} = \frac{\sum_{k=1}^{z} Ei^{t_k}}{zS^{(a)}} 100\% \qquad (21).$$

Thus, at the end of primary crystallization, we know the mean chemical composition of the austenite phase, as well as the quantitative and qualitative composition of the carbide phases in different zones of the formed metal. Equations (11) and (13)-(22) comprise a phenomenological model of the primary non-equilibrium crystallization of the weld pool and the formation of the weld metal. At the end of primary crystallization, we have a weld deposit of complex phase and structural composition that consists of primary carbides and of austenite phases.

### 5. Physicochemical analysis of secondary crystallization.

Secondary crystallization is accompanied by diffusion-controlled evening of the composition of the crystallized metal to the composition specified by expressions (21) and (22), and partial coagulation of the primary carbides along their grain boundaries during cooling. When the temperature for the limiting solubility of carbon and the alloying elements in austenite is reached, the isothermal decomposition of austenite occurs, and the distribution of carbon between the carbide phases is proportional to the CFA of the respective transition element.

The evolution of the system in this stage could be predicted theoretically on the basis of the corresponding phase diagrams. However, the construction of such phase diagrams is extremely difficult for the multi-component alloys under consideration, in which the concentrations of the alloying elements can vary from several percent to tens of percent by weight. Taking into account the features of the crystallization of a weld metal noted above, we should be able to predict its phase constitution on the basis of pseudo-binary phase diagrams on the level of a qualitative estimate. There will still be a probability of a high degree of deviation from reality. This is because equilibrium phase diagrams do not take into account the real nature of the crystallization of a weld metal and the effects of the thermal-straining cycle during high temperature processes, such as welding, as well as the cyclic nature of primary crystallization, which results in chemical non-uniformity of the crystallizing metal. More than 70 years, a similar problem has been solved for certain types of molten metals in a weld pool using the phenomenological Schaeffler constitution diagram [21]. The Schaeffler diagram is a real empirical diagram that is constructed for the weld metal in the initial state after welding for ordinary averaged manual arc welding regimes are shown in Fig. 3.

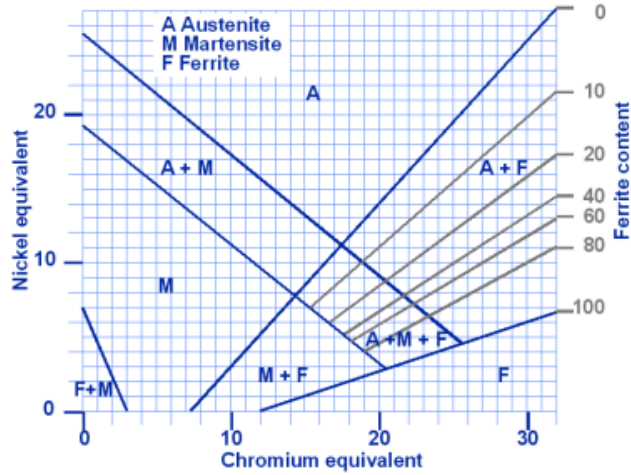

**Fig. 3.** Schaeffler diagram.

Schaeffler diagram indicates a real metal microstructure formed after secondary crystallization for high alloy steel welds but can be applied for a high variety of high temperature processes in steel making processes. The microstructure composition consists of the ferrite, austenite and martensite phases can be calculated by the following equations which are called equivalents. Chromium equivalent is calculated using the weight percentage of ferrite stabilizing elements:

$$Cr_{eq} = \%Cr + \%Mo + 1.5 \cdot \%Si + 0.5 \cdot \%Nb + 0.8 \cdot \%V + 4 \cdot \%Ti \quad (22),$$

and nickel equivalent is calculated using the weight percentage of austenite stabilizing elements:

$$Ni_{eq} = \%Ni + 30 \cdot \%C + 0.5 \cdot \%Mn + 1.6 \cdot Al + 19 \cdot N + 0.3 \cdot Cu \quad (23).$$

However, when a metal contains a considerable amount of carbon (more than 0.12 wt. %) and ferrite-forming elements, are also carbides (NbC, TiC, VC, *etc.*) form in the metal microstructure. The possibility of their formation must be taken into account because the equivalent values of chromium and nickel can deviate significantly in this case from the values calculated using equations (22) and (23) proposed by Schaeffler without consideration of the formation of carbides phases.

The decomposition of austenite begins at 1100-1200 K and is accompanied by the precipitation of secondary carbides, which form mainly with chromium and iron (carbides with the general formulas $Me_3C$, $Me_{23}C_6$, $Me_7C_3$, and $Me_6C$). Empirical relations for predicting the type of carbide formed were determined using literature data [28, 29] and the results of our own research 4-7, 9] on the basis of the ratio between the atomic concentrations of the carbide-forming element and carbon in austenite and the parameters of the thermal-straining cycle during welding. In analogy to (17), we can write:

$$Ei^{(c)} = \sum_{j=1}^{k} w_j \cdot \eta_i \cdot E_C^{(d)} \frac{xA_i}{yA_C} \quad (24).$$

Here $w_j$ is the fraction of carbon in the *j-th* carbide phase relative to the total amount of carbon used to form carbides of the *i-th* alloying element, $E_C^{(d)}$ is the concentration of carbon in the austenite decomposition products, and $\eta_i$ is the coefficient defined by (16). Then the concentrations (wt.%) of the carbide-forming element dissolved in the matrix are given by the expression:

$$Ei^{(b)} = Ei^{(a)} - Ei^{(c)} \qquad (25),$$

and of the carbon dissolved in the matrix are given by the expression:

$$E_C^{(b)} = E_C^{(a)} - \sum_{i=1}^{n}\sum_{j=1}^{k} w_j \cdot \eta_i \cdot Ei^{(c)} \frac{yA_C}{xA_i} \qquad (26)$$

The concentration (wt.%) of the carbide phases formed as a result of secondary crystallization is:

$$Q^{II} = E_C^{(a)} - E_C^{(b)} + \sum_{i=1}^{n} Ei^{(c)} \qquad (27),$$

and the total concentration (wt.%) of the carbide phases in the weld metal is:

$$Q_{Hd} = Q^I + Q^{II} \qquad (28).$$

The concentration (wt.%) of the matrix in the weld deposit is determined from the expression:

$$S^{(b)} = 100\% - Q_{Hd} \qquad (29).$$

The concentrations (wt.%) of carbon and in the matrix is:

$$E_C^b = \frac{E_C^{(b)}}{S^{(b)}} \cdot 100\% \qquad (30),$$

and of the alloying elements in the matrix is:

$$Ei^b = \frac{Ei^{(b)}}{S^{(b)}} 100\% \qquad (31)$$

Equations (24)-(31) comprise a phenomenological model of the secondary crystallization process in the weld metal which enables us to predict the phase constitution of the weld deposit.

## 6. Technological experiments using modeling approach.

The task of the following step is developing of a new flux-cored wire for forming of build-up layer with shock-abrasion resistance properties. The inverse problem of flux-cored wire computation has been solved (using mentioned model) which can provide us with the required chemical composition of build-up metal and as the result it can give us the required mechanical properties. These required properties are achieved by austenite-martensite matrix structure with 10 wt. % of carbides uniformly distributed in it.

Austenite has a FCC structure allows holding a high proportion of carbon in its solution. In our case austenite is used for shock resistance thanks to its impact energy absorbance ability. Martensite has a BCT structure where the carbon atoms constitute a supersaturated solid solution and as a result it has the hardest and strongest properties. Therefore martensite serves as abrasion resistant according to its mechanical properties. The stable carbide phase brings the better toughness and additional abrasion resistance and also ensures uniform distribution of the hardness properties. Under intensive impact loading, some amount of metastable austenite absorbs part of the impact energy and transforms into addition martensite phase.

The mathematical model permits prediction of the composition of the weld metal as a function of the compositions of the starting materials and the technological parameters of the welding process. Prediction of the microstructure of the weld metal is based on computer simulation of a Schaeffler diagram and the process of carbide formation in steels.

A cold-rolled ribbon (1008 steel) was filled with a powder mixture calculated using the mentioned model. The main alloying elements, in final wire, were: graphite, ferrotitanium, chromium and nickel powders. From Hume-Rothery rules it is known that the crystal structures of the solute and the solvent must be the same. Here the mentioned alloying elements should be dissolved in FCC structure (austenite phase). It is also known that the size difference between solute and solvent must be $< \sim 15\%$.

Austenite and carbide are the only phases crystallize during primary crystallization process. Chromium and nickel dissolves well in the austenite formed matrix. However, titanium, because of its high difference in atomic radii as compared to γ-iron, and because of its different lattice structure, poorly dissolves in austenite. Some amount of non dissolved in austenite chromium and titanium forms carbides.

By the end of the secondary crystallization, the stable carbide phases will be stay and residual amount of the alloying metals will be dissolved in the metal matrix.

The required properties for shock-abrasion resistance were the input, and the output are the needed alloying elements and their wt.% of the flux and wt.% of the final wire. The output is presented in Table 1 which presents chemical composition of the base metal (A516), wire band (cold-rolled ribbon, 1008 steel) and alloying elements the band was filled with.

| Component | Density (g/cm$^3$) | Core composition (quantity in 100 kg of FCE, kg) | Relation in the dray mixture of the flux (%) |
|---|---|---|---|
| FeCrC | 3.480 | 11.170 | 41.860 |
| FeTiC | 3.150 | 8.130 | 30.490 |
| Ni powder | 2.960 | 4.470 | 16.740 |
| CaF$_2$ | 1.390 | 2.910 | 10.920 |

| | Composition of the base materials and the build-up layers, wt% | | | | | | | |
|---|---|---|---|---|---|---|---|---|
| | C | Si | Mn | Cr | Mo | Ti | Fe | Ni |
| A 516 | 0.280 | 0.300 | 1.000 | 0.000 | 0.000 | 0.000 | 98.350 | 0.001 |
| Electrode bend | 0.080 | 0.030 | 0.500 | 0.120 | 0.100 | 0.010 | 98.820 | 0.250 |

| Required weld | 1.000 | 0.600 | 0.800 | 5.500 | 0.000 | 2.000 | 86.568 | 3.500 |
| --- | --- | --- | --- | --- | --- | --- | --- | --- |
| Layer 1 | 0.960 | 0.481 | 0.543 | 5.304 | 0.099 | 2.004 | 86.324 | 3.483 |
| Layer 2 | 1.141 | 0.529 | 0.421 | 6.713 | 0.125 | 2.536 | 83.129 | 4.408 |
| Layer 3 | 1.189 | 0.542 | 0.389 | 7.088 | 0.132 | 2.680 | 82.279 | 4.654 |

**Table 1.** Flux cored wire for shock-abrasion resistance output.

Build-up samples were produced using flux cored wire with diameter 1.7 mm by process which was performed by welding machine Kemppi FU 30, PS 3500. The samples were prepared by 3 layers build-up metal. The technological parameters of the process were:
- Current: 250A.
- Voltage: 35V.
- Feed speed: 180 m/h.
- Travel speed: 30 m/h.
- Polarity: Reverse.

The samples were tested on home-made shock-abrasion resistance measuring device. The samples were subjects of the mechanic impact loading with the simultaneous continuous sea sand strew (85 g/min) of the 60 Mesh.

A 12.5 mm thickness sheet of low carbon steel, A516, consisting of 73% ferrite and 27% pearlite phase microstructure, was used as the base metal. The obtained surface (3-rd layer build-up metal) consists from martensite and austenite mixed phase microstructure and carbide stable phase, investigated by Scanning Electron Microscopy (SEM) and shown on Fig. 4.

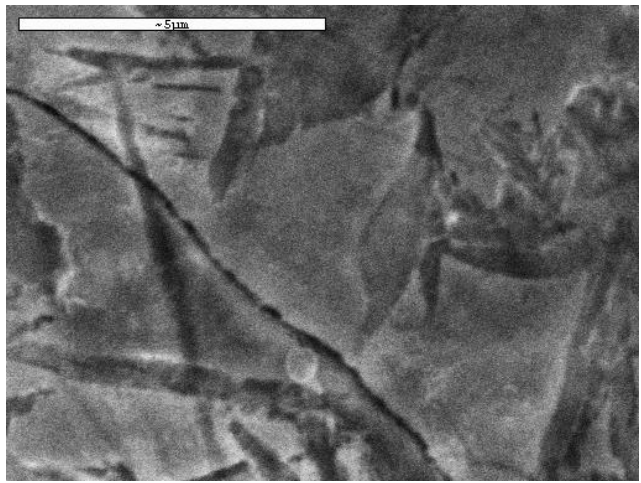

**Fig. 4.** SEM image of surface build-up metal microstructure.

Chemical analysis of the matrix was made with Energy Dispersive X-ray Analysis (EDS) and the received results are shown in Table 2 and compared with the calculated, using the described model, results.

| Element | EDS results (wt.%) | Calculated results (wt.%) |
|---------|-------------------|---------------------------|
| C | 0.97 | 1.19 |
| Si | 0.49 | 0.54 |
| Mn | 0.40 | 0.39 |
| Cr | 7.26 | 7.09 |
| Fe | 84.01 | 82.28 |
| Ni | 3.77 | 4.65 |
| Ti | 3.10 | 2.68 |

**Table 2.** Chemical composition of the weld obtained by EDS analysis and model calculation.

A little differences as seen in Table 2, caused as the result of technique limitation. The detected chemical elements must be rounded to 100% and this limitation doesn't take into account some chemical inclusions that usually found in steels.

As we see from the presented results, the calculated and the real chemical composition results are closed, that emphasize the correctness of the calculations performed by the model.

Hardness tests were made by Rockwell Hardness Tester. The hardness obtained on the 3-rd layer of build-up metal was 56 HRC as compare to the hardness of the base A516 metal - 90 HRB.

The difference between base metal and build up-metal for shock-abrasion resistance results obtained by shock-abrasion tests using special home-made device are presented in the plot shown in Fig. 5. The results of tested specimens presented as the plot of the weight loss per shocked area as a function of time:

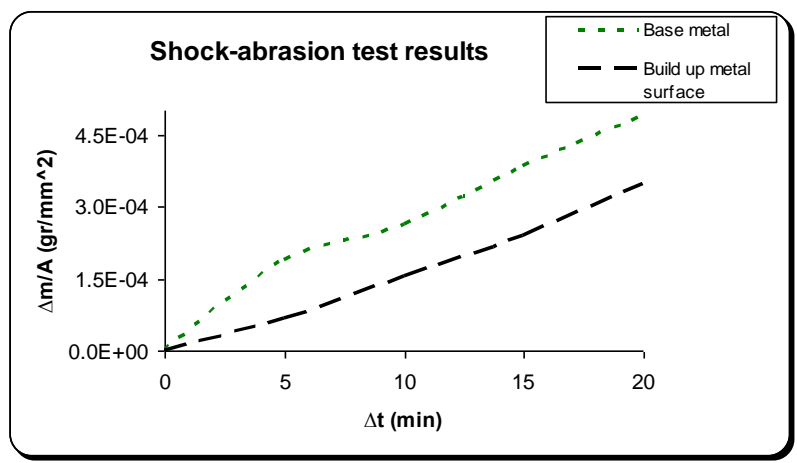

**Fig. 5.** Shock-abrasion test results as a function of time.

It is seen from the plot that the build-up metal has an improved shock-abrasion resistance. It is found that efficiency increased by 29%, what is in a good agreement with the declared aim of the work.

**7. Conclusions.**

A phenomenological model for predicting the chemical composition and structure of the non-equilibrium primary and secondary crystallization which takes into account the carbide phase formation has been developed. This allows the realization of the quantitative prediction of the metal structure and its mechanical properties.

The difficulty which was solved in the purposed model is a kinetic analysis of the interaction of multi-component metallic and slag melts with consideration of the mutual influence of reactions taking place in parallel.

Full-scale testing and investigation have been performed. The obtained results have been analyzed and treated to verify the equivalence of the model. The model was tested for solving the real technological problem, which is developing a new flux-cored wire for forming a build-up layer with shock-abrasion resistance properties. The experimental results confirm the adequacy of the calculations using this model.